\documentclass[11pt]{llncs}
\usepackage{amsmath}
\usepackage{amssymb}
\usepackage{ifpdf}
\usepackage{makeidx}
\usepackage{graphics,graphicx}
\usepackage{psfrag}
\begin{document}
\mainmatter
\newcommand{\pn}{\mathrm{Min}}
\newcommand{\px}{\mathrm{Max}}
\newcommand{\Val}{\mathrm{Val}}
\newcommand{\Width}{\mathit{Width}}
\newcommand{\Value}{\mathit{Value}}
\newcommand{\Ord}{\mathit{Ord}}
\ifpdf
\DeclareGraphicsRule{*}{mps}{*}{}
\else
\DeclareGraphicsRule{*}{eps}{*}{}
\fi
\psfrag{Min}{\normalsize Min}
\psfrag{Max}{\normalsize Max}
\psfrag{start}{\normalsize \em start}
\psfrag{s}{\normalsize \em s}
\psfrag{-1}{\normalsize $-1$}
\psfrag{1}{\normalsize $1$}
\psfrag{2}{\normalsize $2$}
\psfrag{3}{\normalsize $3$}
\psfrag{4}{\normalsize $4$}
\psfrag{5}{\normalsize $5$}
\psfrag{6}{\normalsize $6$}
\psfrag{7}{\normalsize $7$}
\psfrag{8}{\normalsize $8$}
\psfrag{9}{\normalsize $9$}

\title{Solving Min-Max Problems\\with Applications to Games}
\titlerunning{Solving Min-Max Problems with Applications to Games}

\author{Daniel Andersson\thanks{Research supported by \emph{Center for Algorithmic Game Theory}, funded by The Carlsberg Foundation.}}

\authorrunning{Daniel Andersson}
\tocauthor{Daniel Andersson (University of Aarhus)}

\institute{
Department of Computer Science, 
University of Aarhus, 
Denmark\\
\email{koda@daimi.au.dk}}

\maketitle
\begin{abstract}

We refine existing general network optimization techniques, give new characterizations for the class of problems to which they can be applied, and show that they can also be used to solve various two-player games in almost linear time. Among these is a new variant of the network
interdiction problem, where the interdictor wants to destroy high-capacity paths from the
source to the destination using a vertex-wise limited budget of arc removals. We also show that replacing the limit average in mean payoff games by the maximum weight results in a class of games amenable to these techniques. 

\end{abstract} 


\section{Min-Max Problems}

We consider problems whose instances have two parts: a finite discrete part (e.g., a graph), the \emph{structure}, and a list of $n$ objects from some totally ordered set $\mathbb{K}$ (e.g., real numbers), the \emph{comparables}. We adopt the standard random access machine model augmented with the capability of comparing any two of the given comparables in constant time. We will only consider problems where the solution is always one of the comparables in the problem instance (in our machine model, this is output in the form of an index between $1$ and $n$).

A \emph{min-max problem} is such a problem where for any fixed structure $A$ and number of comparables $n$, the resulting function $f_{A,n} : \mathbb{K}^n \to \mathbb{K}$ can be computed by a min-max circuit.
The \emph{ordered version} of a min-max problem is obtained by always supplying as additional input a permutation that non-decreasingly orders the list of comparables. Clearly, if the ordered version can be solved in time $\xi(A,n)$, then by completely sorting the comparables, the original problem can be solved in $O(n \log n + \xi(A,n))$ time and $O(n \log n)$ comparisons. However, it turns out that when $n\log n$ dominates over $\xi(A,n)$, it is possible to improve both of these bounds simultaneously. In particular, we have the following results.

\begin{theorem}
\label{thm:comp}
If the ordered version of a min-max problem with structure $A$ and $n$ comparables
can be solved in time $\xi(A,n) \geq n$, then
the original problem
can be solved in $O(\xi(A,n)\log^* n)$ time and $O(n)$ comparisons.
\end{theorem}

\begin{theorem}
\label{thm:time}
If the ordered version of a min-max problem with structure $A$ and $n$ comparables
can be solved in time $\xi(A,n) \geq n$, then
the original problem
can be solved in $O(\xi(A,n)(1 + \log^* \xi(A,n) - \log^* (\xi(A,n)/n)))$ time.
\end{theorem}

The algorithms are presented in Section \ref{sec:algorithms}. The basic technique employed was used by Gabow and Tarjan \cite{GabowTarjan:1988} to find a bottleneck spanning tree in a digraph with $n$ vertices and $m$ edges in $O(m \log^* n)$ time. Punnen \cite{Punnen:1996} gave a general formulation of the technique and used it to solve the bottleneck Steiner aborescence problem within the same bound.

In this paper, we provide a more detailed analysis, showing that the almost linear running time can be achieved while still maintaining the asymptotically optimal linear number of comparisons (Theorem \ref{thm:comp}), and we maintain the distinction between the structure and the number of comparables to obtain a bound that is never worse than the sorting method (Theorem \ref{thm:time}).

Also, recent work by Litman et al.\cite{LitmanEvenLevi:2007} shows that the functions computable by min-max circuits are exactly the \emph{continuous order statistics} of Rice \cite{Rice:1998} --- continuous\footnote{in the order topology for $\mathbb{K}$ and the box topology based on it for $\mathbb{K}^n$} functions whose output is always one of their inputs --- by showing that they are both the set of functions that commute with every monotone function. This yields two additional characterizations of the class of min-max problems.

While previous work has been focused on network optimization problems, we demonstrate in Section \ref{sec:applications} that the methods can be applied to different classes of two-player games as well.

\section{Algorithms}
\label{sec:algorithms}

We are given an instance of a min-max problem with structure $A$ and comparables $x_1,\ldots,x_n$. To simplify the presentation, we shall assume that all the given $x_i$ are distinct. Suppose that we have an algorithm $\Ord(A,(x'_1,\ldots,x'_n),I)$ that solves the instance $(A,(x'_1,\ldots,x'_n))$ in time $\xi(A,n) \geq n$ when $x'_{I[1]} \leq \cdots \leq x'_{I[n]}$.

The idea is to partition the comparables into groups to obtain a coarse ordered instance of the problem. Due to continuity, solving the coarse instance correctly identifies the group in which the answer to the original instance can be found, and we can recursively continue the search within this group. 

We first consider division into only two groups in each iteration:

\begin{enumerate}
\item $I := \langle 1,2,3,\ldots,n\rangle$ 
\item $lo := 1$
\item $hi := n$
\item While $lo < hi$
\begin{enumerate}
\item $m := \lfloor (hi+lo)/2 \rfloor$
\item \label{partition} Rearrange $I$ so that
$$
\max_{j\in I[lo\ldots m]} x_j < \min_{j\in I[m+1\ldots hi]} x_j.
$$
\item For $j := 1 \ldots lo-1$ : $x'_{I[j]} := x_{I[1]}$.
\item For $j := lo \ldots m$ : $x'_{I[j]} := x_{I[lo]}$.
\item For $j := m+1 \ldots hi$ : $x'_{I[j]} := x_{I[hi]}$.
\item For $j := hi+1 \ldots n$ : $x'_{I[j]} := x_{I[n]}$.
\item If $\Ord(A,(x'_1,\ldots,x'_n),I) \in I[lo\ldots m]$, then $hi := m$.
\item Else $lo := m+1$.

\end{enumerate}
\item Return $lo$.
\end{enumerate}

Let $n_i$ be the value of $hi-lo+1$ in the $i$'th iteration of the while-loop.
Step \ref{partition} can be performed in $O(n_i)$ time and comparisons by a linear time median finding algorithm\cite{BlumFloydPrattRivestTarjan:1972}. Thus, each iteration of the while-loop takes $O(\xi(A,n))$ time, and since $n_i$ is halved ($\pm 1$) with each iteration, the total time is $O(\xi(A,n) \log n)$ and the total number of comparisons is $O(n)$.

An easy way to improve this algorithm is to break the while-loop as soon as we can afford to simply sort the comparables in the remaining interval, i.e., when $n_i \log n_i \leq n$. This brings the number of iterations down to $O(\log\log n)$.

To get even closer to linear time, we need to partition the comparables into more than two groups. Recursive partitioning around the medians (i.e., a prematurely cancelled perfect quicksort) can construct $2^k$ groups in $O(kn_i)$ time. The number of groups will be chosen so as to balance the work spent on partitioning and solving.


If we allot $O(n / i^2)$ time for partitioning in the $i$'th iteration, then we can ensure that $n_{i+1} \leq n_i 2^{-2n/(n_ii^2)}$, and solving this recurrence (see Appendix) shows that the number of iterations drops to $O(\log^* n)$. Since $\sum_{i=1}^{\infty} \frac{1}{i^2}$ converges, the total number of comparisons remains $O(n)$.

If we are willing to give up the $O(n)$ bound on comparisons, we can instead allot $O(\xi(A,n))$ time for partitioning in each iteration and get $O(1 + \log^* \xi(A,n) - \log^* (\xi(A,n)/n))$ iterations (see Appendix for details). Note that if $\xi(A,n) = \Omega(n \log\log\log\cdots \log n)$ for some fixed number of $\log$'s, then this bound is actually $O(1)$.
\section{Applications to Games}
\label{sec:applications}
\subsection{Simple Recursive Games}
Andersson et al.\cite{AnderssonEtAl} introduce the class of \emph{simple recursive games}, which are two-player zero-sum perfect information extensive form games where the game tree is replaced by a game \emph{graph}. Infinite play is interpreted as a zero payoff. Thus, they are similar (but incomparable) to Condon's simple stochastic games \cite{Condon:1992:CSG}, but with no moves of chance and an arbitrary number of different payoffs. Figure 1 shows an example game.

\begin{figure}
\hspace{2em}
\includegraphics[scale=1.15]{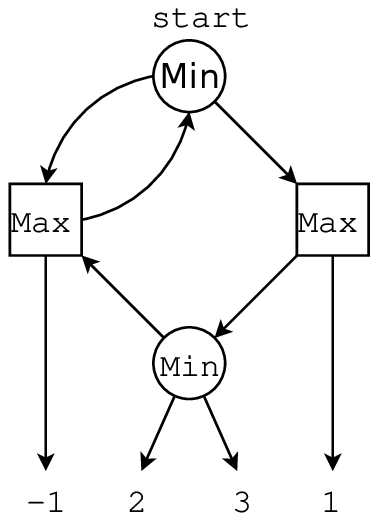}
\hfill
\includegraphics[scale=1.15]{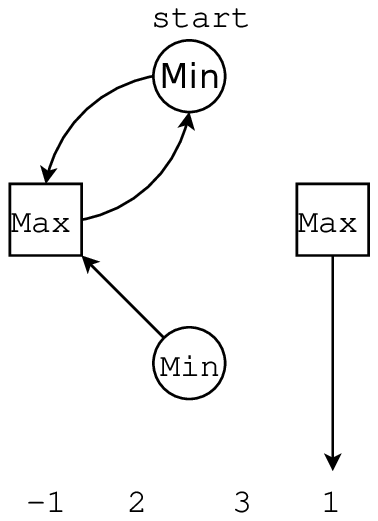}
\hspace{2em}
\caption{A simple recursive game (\emph{left}) and a solution (\emph{right}).}
\end{figure}

In \cite{AnderssonEtAl}, the problem of finding a weak solution (value and minimax strategies for a specified starting position) is reduced to a min-max problem, which is solved using algorithms analogous to those presented herein. However, for the case of optimal number of comparisons, Thorem \ref{thm:comp} constitutes an improvement over the previous result.
\subsection{Maximum Payoff Games}
{\em Mean payoff games} \cite{Zwick:1996:CMPG} is a class of infinite duration games played on a weighted sink-free digraph $(V,E,w)$. Starting from a specified node, two players take turns choosing outgoing arcs to create an infinite path with arcs $e_0,e_1,e_2,\ldots$, and player Min pays to player Max the limit average \begin{equation}\label{eq:mpg}\lim_{n\to\infty} \frac{1}{n}\sum_{0\leq i \leq n} w(e_i).\end{equation}

In {\em discounted payoff games}, (\ref{eq:mpg}) is replaced with a discounted average $\sum_{0\leq i} \lambda^i w(e_i)$, where $\lambda\in [0,1)$ is a parameter. Both classes are closely related to model checking for the modal $\mu$-calculus, and they are interesting from a complexity-theoretic point of view, since they give rise to problems that are among the few natural ones known to be in $\mathrm{NP}\cap\mathrm{coNP}$ but not known to be in P. The best known upper bounds for solving them are randomized subexponential time \cite{BjorklundVorobyov:2007}.

\begin{figure}
\hspace{2em}
\includegraphics[scale=1.15]{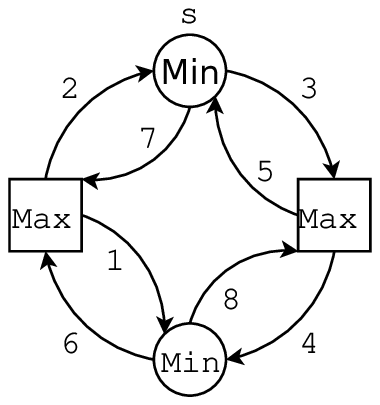}
\hfill
\includegraphics[scale=1.15]{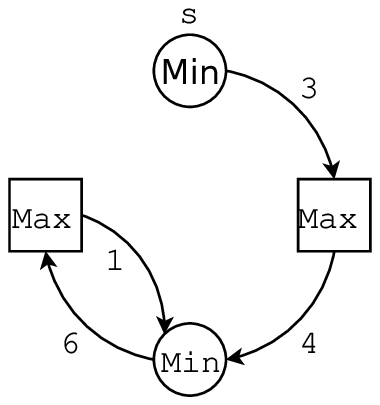}
\hspace{2em}
\caption{A maximum payoff game (\emph{left}) and a solution (\emph{right}).}
\end{figure}

Interestingly, if we change the evaluation function for the stream from limit or discounted average to simply $\max_{0\leq i} w(e_i)$, the complexity of solving the games plummets. Finding the value of a game with a specified starting node $s$ is now a min-max problem, and the ordered version can be solved in linear time as follows:
\begin{enumerate}
\item $\Value[V] := +\infty$
\item While there is an arc
\begin{enumerate}
\item Let $e$ be a max-weight arc, and let $v$ be its tail.
\item If $v$ belongs to Min and $e$ is not its only outgoing arc, remove $e$.
\item Else, increase the weight of $v$'s incoming arcs to $w(e)$, set $\Value[v]:=w(e)$,
remove $v$'s outgoing arcs except $e$, and contract $e$ (its head determining the type of the resulting node).
\end{enumerate}
\item Return $\Value[s]$
\end{enumerate}

\subsection{Widest Path Interdiction}
Motivated by military applications, McMasters and Mustin
\cite{McMastersMustin:1970} defined and studied \emph{network interdiction problems},
where an interdictor with limited resources wants to inhibit the
usefulness of a network.
In particular, {\em shortest path interdiction} is a 
well-studied variant
\cite{IsraelyWood:2002,KhachiyanEtAl:2007}. Phillips
\cite{Philips:1993} considered minimizing the maximum flow.

In this paper we define and consider a new variant: {\em widest path
interdiction}. 
We are given a connected network $(V,E)$ with arc capacities $c : E \to \mathbb{R}^+$, a source $s$, a
destination $t$, and a \emph{budget} $k(v)$ for each vertex $v$. From each 
vertex $v$
the interdictor removes
at most
$k(v)$ outgoing
arcs, so that the width of the widest path from $s$ to $t$ is
minimized (the width of
a path is the
minimum capacity along that path). We could also, as in \cite{KhachiyanEtAl:2007}, allow more general budget constraints,
  specified by a certain class of oracles, without affecting the asymptotic
  running time of our algorithms.

This is a min-max problem.
The ordered version can be solved in $O(|E|)$ time as follows:
\begin{enumerate}
\item $\Width[V] := 0$
\item $\Width[t] := +\infty$
\item While $\mathit{indeg}(t) > 0$
\begin{enumerate}
\item Let $e$ be a max-capacity incoming arc to $t$, and let $v$ be its tail.
\item If $v$'s budget allows it, remove $e$ (this is an
  optimal removal).
\item Else, ensure that arcs to $v$ have capacity at most $\Width[v]:=c(e)$,
remove all outgoing arcs from $v$, and merge $v$
with $t$.
\end{enumerate}
\item Return $\Width[s]$
\end{enumerate}

To implement the algorithm, we need a max-priority queue for the incoming
arcs to $t$. However, the
extracted values form a non-increasing sequence, so we can first replace
the capacities with integers from $1$ to $|E|$ (using the given
permutation) and then use an
array of $|E|$ buckets to stay within linear time.


By Theorem \ref{thm:comp}, the widest path interdiction problem can be solved in
in $O(|E| \log^* |E|)$ time and $O(|E|)$ comparisons, which is also, considering bounds in
$|E|$ only, the best known bound for the (uninterdicted) widest path
problem in directed graphs \cite{GabowTarjan:1988}.

If we instead consider a \emph{global} budget, i.e., any set of at most $k$ arcs may
be removed, then we can solve the widest path interdiction problem by
performing a binary search for the smallest capacity $q$ such that
removing all arcs of capacity at most $q$ yields a network with arc
connectivity at most $k$. Using Dinic's blocking flow algorithm \cite{EvenTarjan:1975} to compute the arc connectivity, the total running time becomes
$O(|E| \min(|E|^{1/2},|V|^{2/3}) \log |E|)$. This is in stark contrast to
shortest path interdiction with a global budget \cite{KhachiyanEtAl:2007}, where the
maximin path length is
NP-hard to approximate within a factor less than $2$.









\subsubsection*{Acknowledgements.}  
The author thanks Peter Bro Miltersen and Troels Bjerre S\o{}rensen for helpful comments and discussions.


\section*{Appendix: Two Recurrences}
\subsection*{$O(n/i^2)$ Time for Partitioning}
First we consider the recurrence
\begin{eqnarray*}
n_1 &=& n\\
n_{i+1} &\leq& n_i 2^{-2n/(n_ii^2)}.
\end{eqnarray*}
Letting $x_i = n/n_i$ we get
\begin{eqnarray*}
x_1 &=& 1\\
x_{i+1} &\geq& x_i 2^{2x_i/i^2}.
\end{eqnarray*}
\begin{lemma}
\label{lem:log}
For $i\geq 4$, $x_i \geq i^2 \log_2 (i+1)$.
\end{lemma}
\begin{proof}
By induction.\qed
\end{proof}
\begin{lemma}
\label{lem:tow1}
For $i\geq 4$, $x_{i+2} \geq 2^{x_i}$.
\end{lemma}
\begin{proof}
By definition and Lemma \ref{lem:log}, we have for $i\geq 4$,
\begin{eqnarray*}
x_{i+2} &\geq& x_{i+1} 2^{\frac{2x_{i+1}}{(i+1)^2}}\\
        &\geq& 2^{\frac{2}{(i+1)^2} x_i 2^{2x_i / i^2}}\\
        &\geq& 2^{x_i}.
\end{eqnarray*}
\qed
\end{proof}
It follows from Lemma \ref{lem:tow1} that $\min\{ i : n_i \leq 1\} = O(\log^* n)$.
\subsection*{$O(\xi(A,n))$ Time for Partitioning}
We let $m = \xi(A,n)$ and consider the recurrence
\begin{eqnarray*}
n_1 &=& n\\
n_{i+1} &\leq& n_i 2^{-m/n_i}.
\end{eqnarray*}
Letting $x_i = n/n_i$ we get
\begin{eqnarray*}
x_1 &=& 1\\
x_{i+1} &\geq& x_i 2^{x_i m/n} \\
        &\geq& 2^{x_1 m/n},
\end{eqnarray*}
and thus $\min\{ i : n_i \leq 1\} = O(\log_{2^{m/n}}^* n) = O(1+ \log^* m - \log^* (m/n))$.

\end{document}